\documentclass{article}
\usepackage[utf8]{inputenc}
\usepackage{graphicx}
\usepackage{hyperref}
\usepackage[table,xcdraw]{xcolor}
\usepackage{adjustbox}
\usepackage{subcaption}
\usepackage{framed}

\begin{document}
\begin{center}

\begin{Large}
\vspace{0.5cm}
\textbf{Ensemble Boost: Greedy Selection for Superior Recommender Systems \\}
\end{Large}
\vspace{0.5cm}

\begin{large}

Zainil Mehta \\
Matriculation Number: 1634574\\
\vspace{0.5cm}
University of Siegen \\
\vspace{0.1cm}
Chair: Intelligent Systems Group (ISG) \\
\vspace{0.1cm}
Supervisor: Prof. Dr.-Ing. Joeran Beel \\
\vspace{0.1cm}
Advisor:  Mr. Tobias Vente\\
\vspace{0.2cm}
March 2024 \\

\end{large}

\end{center}

\begin{abstract}

Ensemble techniques have demonstrated remarkable success in improving predictive performance across various domains by aggregating predictions from multiple models~\cite{inbook}. In the realm of recommender systems, this research explores the application of ensemble technique to enhance recommendation quality. Specifically, we propose a novel approach to combine top-k recommendations from ten diverse recommendation models resulting in superior top-n recommendations using this novel ensemble technique. Our method leverages a Greedy Ensemble Selection(GES) strategy, effectively harnessing the collective intelligence of multiple models. We conduct experiments on five distinct datasets to evaluate the effectiveness of our approach. Evaluation across five folds using the NDCG metric reveals significant improvements in recommendation accuracy across all datasets compared to single best performing model. Furthermore, comprehensive comparisons against existing models underscore the efficacy of our ensemble approach in enhancing recommendation quality. Our ensemble approach yielded an average improvement of 21.67\% across different NDCG@N metrics and the five datasets, compared to single best model. The popularity recommendation model serves as the baseline for comparison. This research contributes to the advancement of ensemble-based recommender systems, offering insights into the potential of combining diverse recommendation strategies to enhance user experience and satisfaction. By presenting a novel approach and demonstrating its superiority over existing methods, we aim to inspire further exploration and innovation in this domain.
\end{abstract}

\section{Introduction}

With the expanding usage of the Internet, individuals face the challenge of information overflow, characterized by an overwhelming surplus of data that often leaves users overwhelmed and unable to effectively utilize the available information to their advantage~\cite{TN_libero_mab2}. Since recommender systems have emerged, it has revolutionized the way users interact with online platforms. These systems have become indispensable tools for navigating through immense volume of information available on the internet, helping users in discovering relevant information tailored to their preferences and needs~\cite{10.1007/s11042-020-09949-5}. These systems, found across diverse platform from e-commerce to streaming services, rely on advanced models to analyze user behavior and provide personalized recommendations~\cite{Ricci2011}. By analyzing user behavior and historical data, recommender systems aim to predict items that users are likely to be interested in, thus facilitating decision-making and enhancing user experience. Over the years, a range of recommendation technique have been developed, each with its own strengths and limitations. 
\\
\\
However, traditional recommender systems face several challenges that hinder their effectiveness in delivering personalized recommendations. These challenges include cold start problems, where new users or items lack sufficient historical data for accurate recommendations, scalability issues when dealing with large datasets, and the inability to effectively capture user preferences in dynamic environments~\cite{10.1145/564376.564421}. These limitations underscore the need for innovative approaches to enhance recommendation quality and address the evolving needs of users.
\\
\\
Ensembling techniques have emerged as powerful method for improving predictive performance across various machine learning domains~\cite{inbook}. It involves combining prediction from various models to produce a more robust and accurate prediction. By leveraging the diversity of different models and learning algorithms, ensembles can often outperform best performing individual models, leading to significant improvement in prediction accuracy and generalization performance. Many researches now exists showing that ensemble learning often increases performance~\cite{4053111}. While ensembling has been studied extensively in classification and regression, its application in recommender system remains relatively unexplored. This research seeks to bridge this gap by investigating the integration of ensemble techniques into recommender systems to enhance recommendation quality.
\\
\\
Therefore, in this paper we investigate the usage of ensemble techniques into recommeder systems, with a goal of not only enhancing the quality of recommendations but also diversity of recommendations offered to users through the utilization of different combinations of recommendation models. An essential component of ensembles is the combination method, which plays a pivotal role in combining predictions from multiple models. This method employs weighted ranking to create the final recommendation list. The method used here is tested on five datasets namely MovieLens-100k, MovieLens-1m, ciaodvd, hetrec-lastfm, and citeulike-a. The ten recommendation models are implicit-matrix factorization, user-knn, item-knn, alternating-least-squares, bayesian-personalized-ranking, logistic-matrix factorization, item-item-cosine, item-item-tfidf, item-item-bm25, and popularity.
\\
\\
The approach utilizes Forward Greedy Ensemble Selection~\cite{Vente2022} to identify the optimal ensemble combination, aiming to surpass the performance of any singular recommendation model employed. Each individual recommendation model, as well as the resulting ensemble, is evaluated and compared on the test dataset using the NDCG(Normalized-Discounted Cummulative Gain) metric. It is a widely adopted metric in the field of information retrieval and recommendation systems~\cite{10.1145/582415.582418}. It quantifies the effectiveness of a ranked list by considering both the relevance of items and their positions in the list. NDCG provides a normalized measure that ranges between 0 and 1, where higher values indicate better alignment with user preferences.
\\
\\
Each model has been fine-tuned for specific top-N\footnote{N refers to the ndcg@N, for which the model is optimized } recommendations. Consequently, the ensemble technique comprises varying top-k\footnote{k refers to top-k recommendation taken into consideration} recommendations across different models, but ultimately considers only the top-N items in the ensemble. The NDCG scores of the ensemble are assessed for different top-k recommendations, revealing performance enhancements across almost all scenarios as the k value increases.
\\
\\
\textbf{The paper is structured as follows: Section 2 provides a review of the related literature concerning Ensembling and Recommender Systems. In Section 3, we detail the methodology, focusing on the approach to ensemble combination. Sections 4 encompass the presentation and discussion of results and experiments, followed by conlusion of the research in Section 5 .}

\section{Related Work}

Ensembling techniques have gained significant attention in various domains for their ability to enhance predictive performance by combining multiple models~\cite{10.1007/s11042-020-09949-5}. Caruana et al.~\cite{4053111} demonstrated the effectiveness of ensemble selection in improving predictive accuracy across diverse datasets. Their work emphasized the importance of selecting an optimal ensemble of models to maximize performance.  Forouzandeh et al.~\cite{10.1007/s11042-020-09949-5} applied ensemble learning to recommender systems, specifically focusing on MovieLens dataset. They utilized graph embedding techniques in conjunction with ensemble learning to improve recommendation quality. While their approach showcased promising results, it primarily focused on a specific dataset, leaving room for exploration across broader datasets and ensemble methods.
\\
\\
Evaluation metrics play a crucial role in assessing the effectiveness of recommendation algorithms and ensembles. Järvelin and Kekäläinen~\cite{10.1145/582415.582418} introduced the Non-Discounted Cumulative Gain (NDCG) metric, which measures the relevance of recommended items in ranked lists. This metric has become widely adopted in evaluating recommendation systems.
\\
\\
In the realm of automated recommender systems, various ensembling techniques have been explored to enhance the performance of recommendation models. One notable approach is presented by Vente et al. ~\cite{Vente2022}~\cite{10.1145/3604915.3610656}, who introduced LensKit-Auto, an experimental toolkit within the LensKit framework. Their work highlights the use of greedy ensemble selection, a method that iteratively selects the best-performing models to create an ensemble that maximizes overall recommendation quality. This approach demonstrates significant improvements in recommendation performance by leveraging the strengths of multiple algorithms, making it highly relevant to our research on ensembling techniques in recommender systems.
\\
\\
While existing literature provides insights into ensembling techniques and their application in recommender systems, our research contributes to this field by proposing a ensembling approach to combine recommendation algorithms and its recommendations using GES. Through extensive experimentation on multiple datasets, we evaluate the effectiveness of our ensemble method in improving recommendation quality and diversity. By leveraging prior research and employing ensemble selection strategies, our work contributes to the ongoing exploration of ensemble-based recommender systems.

\section{Method}
\subsection{Datasets and Dataset Partitioning}

The study utilized five datasets: MovieLens-100k, MovieLens-1m, ciaodvd, hetrec-lastfm, and citeulike-a, each varying in size. A prediction matrix was constructed, containing user-item interactions along with the predictions generated by the various algorithms utilized in the study. Each entry in the matrix represents the prediction score generated by a specific model for a given user. The prediction matrix serves as the foundation for evaluating the performance of individual recommendation algorithms and ensemble technique. 
\\
\\
Each dataset is partitioned in five folds and each fold has its training (60\%),validation(20\%) and testing subset(20\%). The overall performance of each model and ensembles were then averaged across all folds for each dataset. Moreover, the prediction matrices were optimized for NDCG@5, NDCG@10, and NDCG@20 for each dataset. This approach ensured comprehensive assessment across different subsets of the data, minimizing the risk of overfitting.

\subsection{Models}
Ten recommendation models were employed to generate recommendations tailored to the characteristics of each dataset.  These models included Implicit Matrix Factorization(I-MF), User-based knn(U-KNN), Item-based knn(I-KNN), Alternating Least Squares(ALS), Bayesian Personalized Ranking(BPR), Logistic Matrix Factorization(L-MF), Item-Item-Cosine Similarity(I-I-COSINE), Item-Item-tfidf Similarity(I-I-TFIDF), Item-Item-BM25 Similarity(I-I-BM25), and Popularity-based(PPL) Recommender. These models were further used to form ensembles.

\subsection{Ensemble Method: Weighted Ranking Method}
The prediction score of an item assigned by a model estimates the likelihood or relevance of an item to a given user. A higher prediction score indicates a higher rank in the recommendation list for that user. The proposed ensemble technique utilizes a weighted ranking method to combine the predictions of individual model effectively. This method is based on the principle that algorithms that performs well, as well as frequently recommended items by these models, are assigned higher weights in the ensemble.
\\
\\
Each recommender model assigns scores to items for a specific user, with the score range normalized between 0 and 1 for all models. Additionally, each model's performance is measured using NDCG scores on the validation dataset. NDCG score is the ratio of Discounted Cumulative Gain (DCG) and Ideal Discounted Cumulative Gain (IDCG)~\cite{article}. For each user, if the item is present in the validation dataset, it has a numerator of 1 and 0 otherwise in the DCG formula. IDCG is calculated by assuming all items recommended are present in the validation dataset. The NDCG score of the models is the average of the NDCG of individual users. The model's NDCG scores  for each fold are further stored in a dataframe. 

\[
NDCG = \frac{DCG}{IDCG}
\]

Where:
\[
DCG = \sum_{i=1}^{n} \frac{{rel_i}}{\log_2(i+1)}
\]
\[
IDCG = \sum_{i=1}^{n} \frac{1}{\log_2(i+1)}
\]
\\
Here, \(rel_i\) is equal to 1 if the item is in validation dataset and 0 otherwise , and \(n\) denotes the total number of items considered.
\\
\\
To prioritize the contribution of each models in the ensemble, models are arranged in descending order of their NDCG scores, assigning higher weightage to models with higher performance.
\\
\captionsetup{format=plain, justification=centering, labelfont=bf}
\begin{figure}[htbp]
  \centering
  \includegraphics[width=1\textwidth]{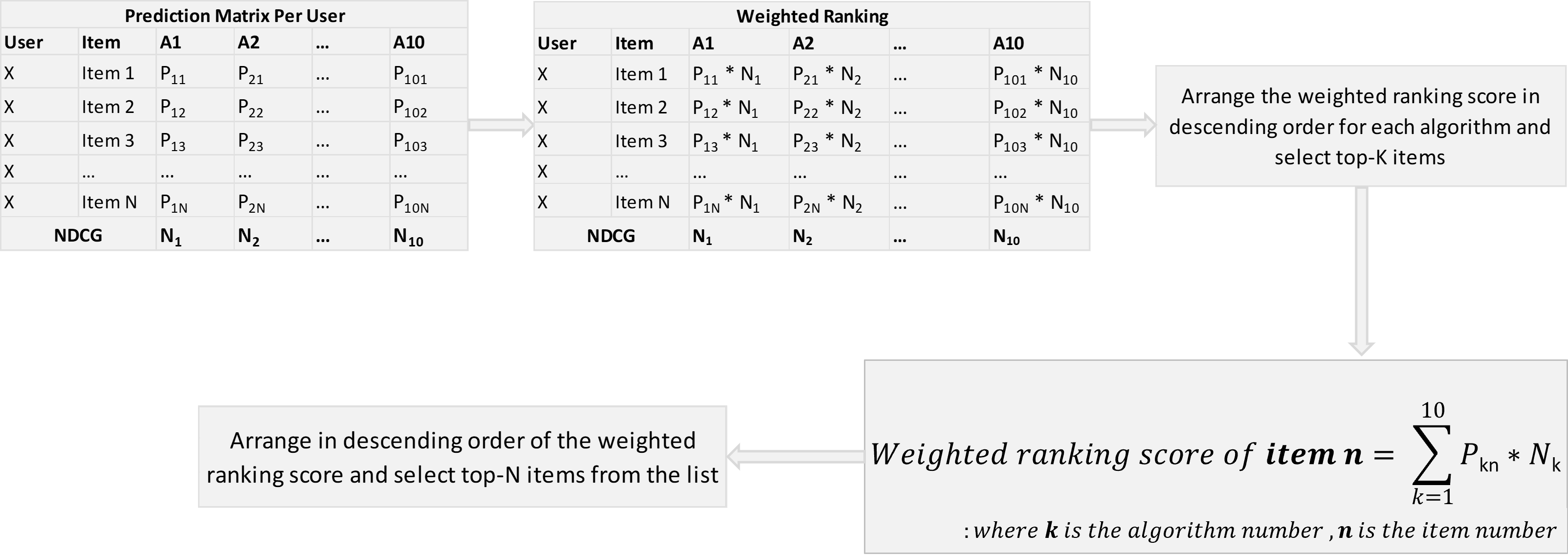}
  \caption{Steps to generate  ensembled list of recommendations of each user}
  \label{fig:ensemble_method}
\end{figure}
\\
For each user and a specific k in the top-k recommendation list, the weighted ranking score is calculated by multiplying the normalized prediction score of each model by its overall NDCG score on the validation subset that is stored in the dataframe. This process is repeated for all ten models. The recommendation list of top k items, for each model is taken into consideration. If an item appears in multiple recommendation lists of different models, its scores are summed up to reflect its collective relevance across models, which result in assigning higher overall scores to items that are consistently recommended across multiple models. All unique items receive scores and are then arranged in descending order of the new weighted ranking score.
\\
\\
Subsequently, only the top-N\footnote{Note:  The algorithms used in ensembling have been optimized for the NDCG@N metric. Considering a larger number of items in the recommendation list of each model may alter the overall ranking in the ensemble list. Therefore, we first compile a list of top-k recommendations and subsequently evaluate only the top-N items for assessment.} items, optimized for specific NDCG@N metrics, are selected to comprise the final ensemble list for each algorithm. (Figure \ref{fig:ensemble_method} shows the overview of the ensembling method)
\\
\\
Next, the ensemble list of recommendations for all users is combined, and the ensemble score is calculated using the NDCG metrics. This novel ensemble technique aims to leverage the strengths of individual algorithms while mitigating their weaknesses, ultimately enhancing the quality and diversity of recommendations provided to users.

\subsection{Selecting best performing ensemble}
\begin{figure}[htbp]
  \centering
  \includegraphics[width=1\textwidth]{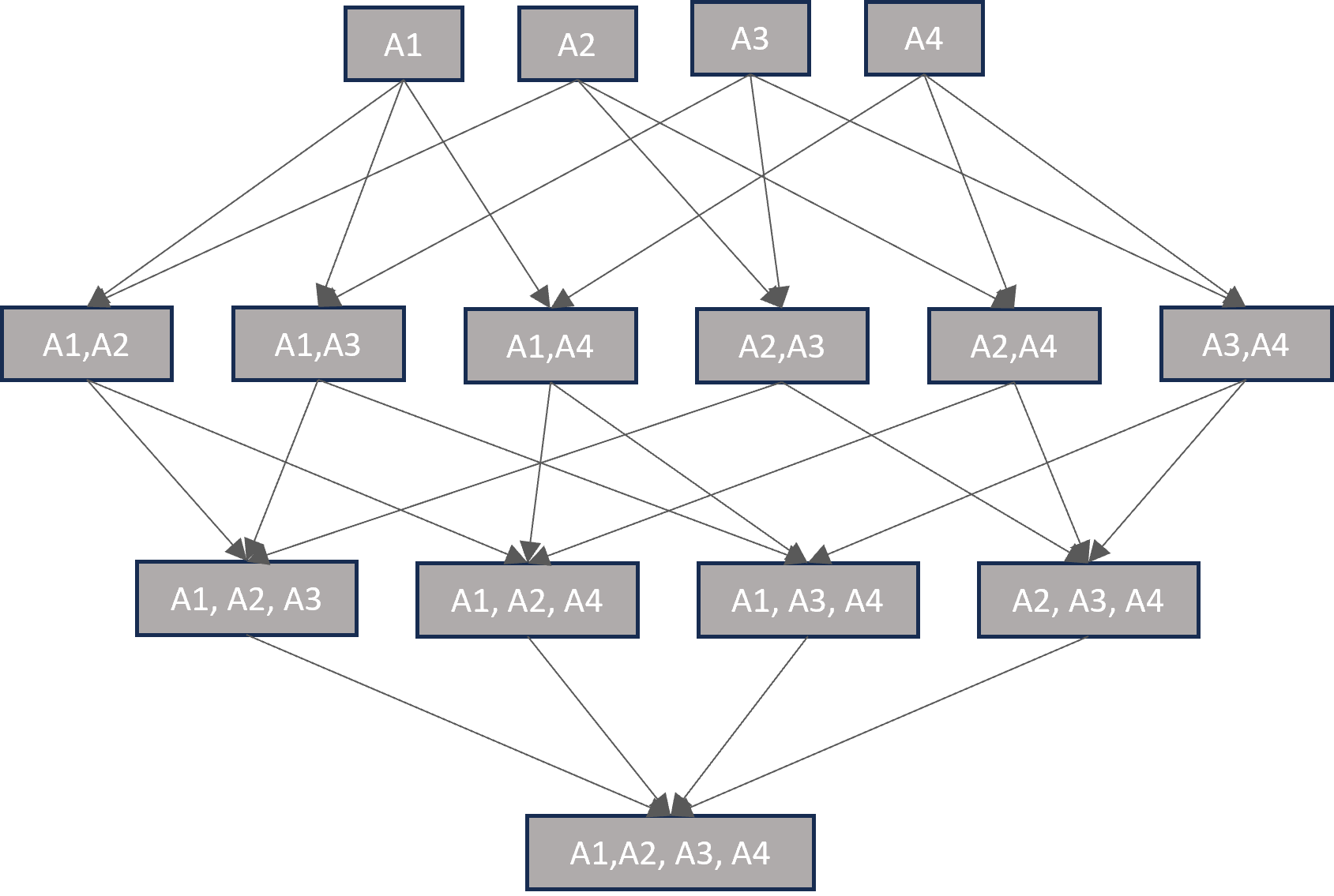}
  \caption{Example of Forward Greedy Ensemble Selection on four models}
  \label{fig:GES}
\end{figure}

Forward Greedy Ensemble Selection (GES) is employed to combine recommendations from the ten models. Initially, pairs of each models are combined, followed by combinations of three, four, five, and so on until all models are covered. Each ensemble is evaluated with the NDCG score using the test subset.
\\
\\
The process of Forward GES is illustrated in Figure \ref{fig:GES}, which provides an example of its application with four different models. This iterative approach allows us to explore a wide range of ensemble combinations systematically.
\\
\captionsetup{format=plain, justification=centering, labelfont=bf}
\\
Throughout our research, we retain all generated ensemble combinations and assess their performance individually, which requires considerable computational resources. This iterative process is repeated across all five folds of the dataset. The final NDCG score on test subset for each model and various ensemble combinations is then averaged across all five folds for a specific dataset. 
\\
\\
The ensemble that yields the highest NDCG score represents the outcome of our ensemble selection method employing Forward Greedy Ensemble Selection. This approach enables us to identify the most effective combination of models for generating recommendation lists with enhanced quality and relevance.

\subsection{Experimental setup}
Ensembling is a post processing step, combining strengths and weakness of different models. The experiments are run on the OMNI cluster of University of Siegen. The code is written in Python and the main libraries used are pandas, numpy, and sklearn. A JSON file is used to configure the settings of the experiment. S-Batch script is used to submit jobs to the cluster. The code is hosted on the 
\href{https://github.com/ZainilMehta/Ensembling_of_RecSys}{github}.

\section{Results and Discussion}
As mentioned before the prediction matrices used for ensebmling have been optimized for N=5, 10, and 20 in NDCG@N. Here, we undertake a comparative analysis focusing on the NDCG scores of the top ensemble model against those of the best-performing individual model. However, it's imperative to present results across all models, as their performance may vary across different datasets. The experimental findings yeilds the following results upon applying the ensemble method across diverse datasets:
\\
\begin{table}[ht]
\centering
\begin{adjustbox}{width=1\textwidth}
\begin{tabular}{llllllll}
\hline
 &
  \cellcolor[HTML]{F2F2F2}\textbf{ciaodvd} &
  \cellcolor[HTML]{F2F2F2}\textbf{citeulike-a} &
  \cellcolor[HTML]{F2F2F2}\textbf{hetrec-lastfm} &
  \cellcolor[HTML]{F2F2F2}\textbf{ml-1m} &
  \cellcolor[HTML]{F2F2F2}\textbf{ml-100k} &
  \cellcolor[HTML]{F2F2F2}\textbf{across-all-datasets} &
  \multicolumn{1}{c}{\cellcolor[HTML]{F2F2F2}\textbf{\% relative to PPL}} \\ \hline
\cellcolor[HTML]{F2F2F2}\textbf{ALS} &
  0.023 &
  0.076 &
  0.175 &
  0.256 &
  \cellcolor[HTML]{B4C6E7}0.253 &
  0.1566 &
  88\% \\
\cellcolor[HTML]{F2F2F2}\textbf{BPR} &
  0.013 &
  0.064 &
  0.099 &
  0.11 &
  0.146 &
  0.0864 &
  4\% \\
\cellcolor[HTML]{F2F2F2}\textbf{I-MF} &
  0.025 &
  0.126 &
  0.184 &
  0.203 &
  0.193 &
  0.1462 &
  76\% \\
\cellcolor[HTML]{F2F2F2}\textbf{I-I-BM25} &
  0.027 &
  0.124 &
  0.199 &
  0.256 &
  0.246 &
  0.1704 &
  105\% \\
\cellcolor[HTML]{F2F2F2}\textbf{I-I-COSINE} &
  0.012 &
  0.095 &
  0.201 &
  0.236 &
  0.229 &
  0.1546 &
  86\% \\
\cellcolor[HTML]{F2F2F2}\textbf{I-I-TFIDF} &
  0.019 &
  0.109 &
  0.201 &
  0.247 &
  0.236 &
  0.1624 &
  95\% \\
\cellcolor[HTML]{F2F2F2}\textbf{I-KNN} &
  0.015 &
  0.11 &
  \cellcolor[HTML]{B4C6E7}0.207 &
  0.24 &
  0.236 &
  0.1616 &
  94\% \\
\cellcolor[HTML]{F2F2F2}\textbf{L-MF} &
  0.02 &
  0.064 &
  0.155 &
  0.171 &
  0.205 &
  0.123 &
  48\% \\
\cellcolor[HTML]{F2F2F2}\textbf{PPL} &
  0.019 &
  0.01 &
  0.079 &
  0.155 &
  0.153 &
  0.0832 &
  0\% \\
\cellcolor[HTML]{F2F2F2}\textbf{U-KNN} &
  \cellcolor[HTML]{B4C6E7}0.03 &
  \cellcolor[HTML]{B4C6E7}0.13 &
  0.184 &
  \cellcolor[HTML]{B4C6E7}0.258 &
  0.25 &
  0.1704 &
  105\% \\
\rowcolor[HTML]{C6E0B4} 
\cellcolor[HTML]{F2F2F2}\textbf{Ensemble-Method} &
  0.036 &
  0.136 &
  0.215 &
  0.269 &
  0.271 &
  0.1854 &
  123\% \\ \hline
\end{tabular}
\end{adjustbox}
\caption{Comparison of NDCG@5 scores of Individual model and Ensemble Method across various datasets}
\label{NDCG@5}
\end{table}
\begin{table}[ht]
\centering
\begin{adjustbox}{width=1\textwidth}
\begin{tabular}{llllllll}
\hline
 &
  \cellcolor[HTML]{F2F2F2}\textbf{ciaodvd} &
  \cellcolor[HTML]{F2F2F2}\textbf{citeulike-a} &
  \cellcolor[HTML]{F2F2F2}\textbf{hetrec-lastfm} &
  \cellcolor[HTML]{F2F2F2}\textbf{ml-1m} &
  \cellcolor[HTML]{F2F2F2}\textbf{ml-100k} &
  \cellcolor[HTML]{F2F2F2}\textbf{across-all-datasets} &
  \cellcolor[HTML]{F2F2F2}\textbf{\% relative to popularity} \\ \hline
\cellcolor[HTML]{F2F2F2}\textbf{ALS} &
  0.02 &
  0.066 &
  0.147 &
  0.234 &
  \cellcolor[HTML]{B4C6E7}0.232 &
  0.1398 &
  84\% \\
\cellcolor[HTML]{F2F2F2}\textbf{BPR} &
  0.013 &
  0.027 &
  0.082 &
  0.119 &
  0.173 &
  0.0828 &
  9\% \\
\cellcolor[HTML]{F2F2F2}\textbf{I-MF} &
  0.022 &
  0.109 &
  0.159 &
  0.189 &
  0.184 &
  0.1326 &
  75\% \\
\cellcolor[HTML]{F2F2F2}\textbf{I-I-BM25} &
  0.024 &
  0.106 &
  0.168 &
  0.234 &
  0.221 &
  0.1506 &
  99\% \\
\cellcolor[HTML]{F2F2F2}\textbf{I-I-COSINE} &
  0.01 &
  0.082 &
  0.169 &
  0.217 &
  0.21 &
  0.1376 &
  82\% \\
\cellcolor[HTML]{F2F2F2}\textbf{I-I-TFIDF} &
  0.016 &
  0.094 &
  0.168 &
  0.225 &
  0.218 &
  0.1442 &
  90\% \\
\cellcolor[HTML]{F2F2F2}\textbf{I-KNN} &
  0.013 &
  0.096 &
  \cellcolor[HTML]{B4C6E7}0.175 &
  0.216 &
  0.212 &
  0.1424 &
  88\% \\
\cellcolor[HTML]{F2F2F2}\textbf{L-MF} &
  0.018 &
  0.058 &
  0.135 &
  0.161 &
  0.191 &
  0.1126 &
  49\% \\
\cellcolor[HTML]{F2F2F2}\textbf{PPL} &
  0.016 &
  0.009 &
  0.07 &
  0.142 &
  0.142 &
  0.0758 &
  0\% \\
\cellcolor[HTML]{F2F2F2}\textbf{U-KNN} &
  \cellcolor[HTML]{B4C6E7}0.026 &
  \cellcolor[HTML]{B4C6E7}0.112 &
  0.157 &
  \cellcolor[HTML]{B4C6E7}0.235 &
  0.225 &
  0.151 &
  99\% \\
\rowcolor[HTML]{C6E0B4} 
\cellcolor[HTML]{F2F2F2}\textbf{Ensemble-Method} &
  0.03 &
  0.117 &
  0.183 &
  0.247 &
  0.243 &
  0.164 &
  116\% \\ \hline
\end{tabular}
\end{adjustbox}
\caption{Comparison of NDCG@10 scores of Individual model and Ensemble Method across various datasets}
\label{NDCG@10}
\end{table}

\begin{table}[ht]
\centering
\begin{adjustbox}{width=1\textwidth}
\begin{tabular}{llllllll}
\hline
 &
  \cellcolor[HTML]{F2F2F2}\textbf{ciaodvd} &
  \cellcolor[HTML]{F2F2F2}\textbf{citeulike-a} &
  \cellcolor[HTML]{F2F2F2}\textbf{hetrec-lastfm} &
  \cellcolor[HTML]{F2F2F2}\textbf{ml-1m} &
  \cellcolor[HTML]{F2F2F2}\textbf{ml-100k} &
  \cellcolor[HTML]{F2F2F2}\textbf{across-all-datasets} &
  \cellcolor[HTML]{F2F2F2}\textbf{\% relative to PPL} \\ \hline
\cellcolor[HTML]{F2F2F2}\textbf{ALS} &
  0.016 &
  0.052 &
  0.108 &
  \cellcolor[HTML]{B4C6E7}0.199 &
  \cellcolor[HTML]{B4C6E7}0.193 &
  0.1136 &
  79\% \\
\cellcolor[HTML]{F2F2F2}\textbf{BPR} &
  0.012 &
  0.026 &
  0.078 &
  0.103 &
  0.125 &
  0.0688 &
  9\% \\
\cellcolor[HTML]{F2F2F2}\textbf{I-MF} &
  0.018 &
  \cellcolor[HTML]{B4C6E7}0.085 &
  0.123 &
  0.167 &
  0.161 &
  0.1108 &
  75\% \\
\cellcolor[HTML]{F2F2F2}\textbf{I-I-BM25} &
  \cellcolor[HTML]{B4C6E7}0.02 &
  0.08 &
  0.123 &
  0.195 &
  0.186 &
  0.1208 &
  91\% \\
\cellcolor[HTML]{F2F2F2}\textbf{I-I-COSINE} &
  0.008 &
  0.063 &
  0.123 &
  0.182 &
  0.173 &
  0.1098 &
  73\% \\
\cellcolor[HTML]{F2F2F2}\textbf{I-I-TFIDF} &
  0.013 &
  0.071 &
  0.124 &
  0.188 &
  0.182 &
  0.1156 &
  82\% \\
\cellcolor[HTML]{F2F2F2}\textbf{I-KNN} &
  0.011 &
  0.073 &
  \cellcolor[HTML]{B4C6E7}0.128 &
  0.176 &
  0.172 &
  0.112 &
  77\% \\
\cellcolor[HTML]{F2F2F2}\textbf{L-MF} &
  0.015 &
  0.045 &
  0.102 &
  0.143 &
  0.164 &
  0.0938 &
  48\% \\
\cellcolor[HTML]{F2F2F2}\textbf{PPL} &
  0.013 &
  0.007 &
  0.054 &
  0.123 &
  0.12 &
  0.0634 &
  0\% \\
\cellcolor[HTML]{F2F2F2}\textbf{U-KNN} &
  \cellcolor[HTML]{B4C6E7}0.02 &
  \cellcolor[HTML]{B4C6E7}0.084 &
  0.115 &
  \cellcolor[HTML]{B4C6E7}0.195 &
  0.187 &
  0.1202 &
  90\% \\
\rowcolor[HTML]{C6E0B4} 
\cellcolor[HTML]{F2F2F2}\textbf{Ensemble-Method} &
  0.024 &
  0.097 &
  0.148 &
  0.219 &
  0.211 &
  0.1398 &
  121\% \\ \hline
\end{tabular}
\end{adjustbox}
\caption{Comparison of NDCG@20 scores of Individual model and Ensemble Method across various datasets}
\label{NDCG@20}
\end{table}

\begin{figure}[tbp]
  \begin{subfigure}{0.6\textwidth}
    \includegraphics[width=\linewidth]{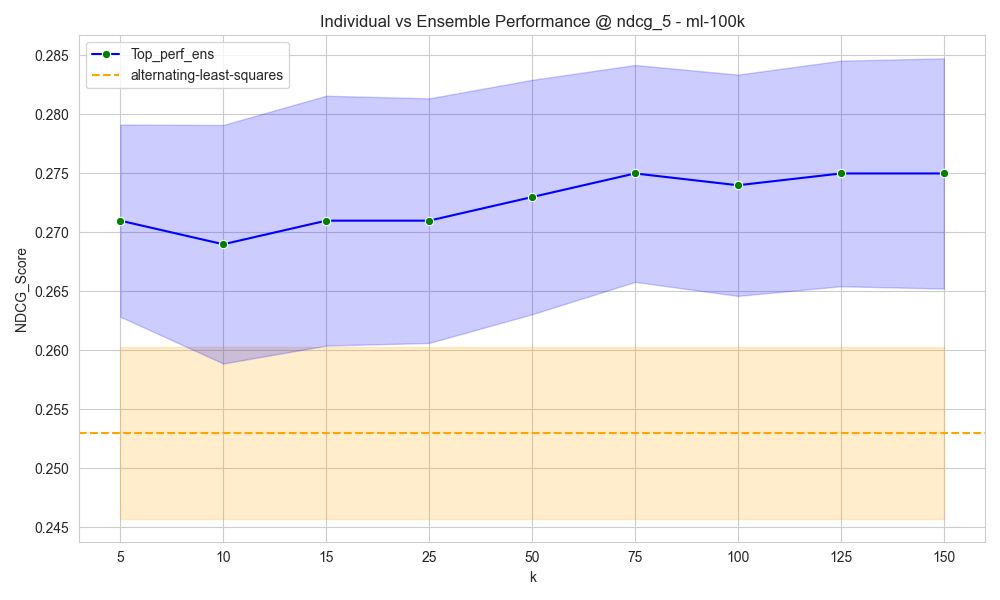} 
    \captionsetup{font=small}
    \caption{ml-100k}
    \label{fig:5_ml-100k}
  \end{subfigure}
  \hfill
  \begin{subfigure}{0.6\textwidth}
    \includegraphics[width=\linewidth]{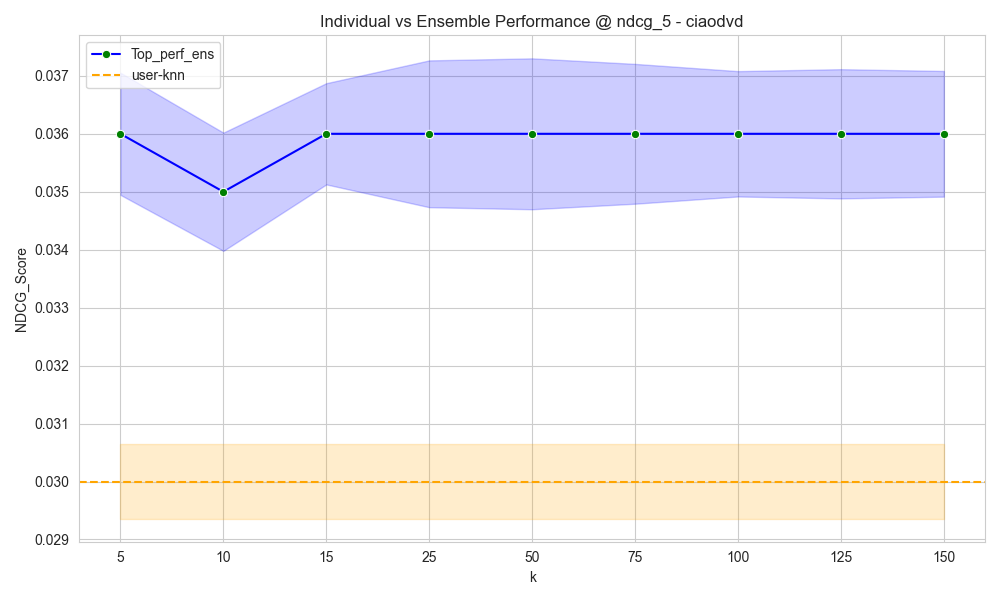} 
    \captionsetup{font=small}
    \caption{ciaodvd}
    \label{fig:5_ciaodvd}
  \end{subfigure}
  \begin{subfigure}{0.6\textwidth}
    \includegraphics[width=\linewidth]{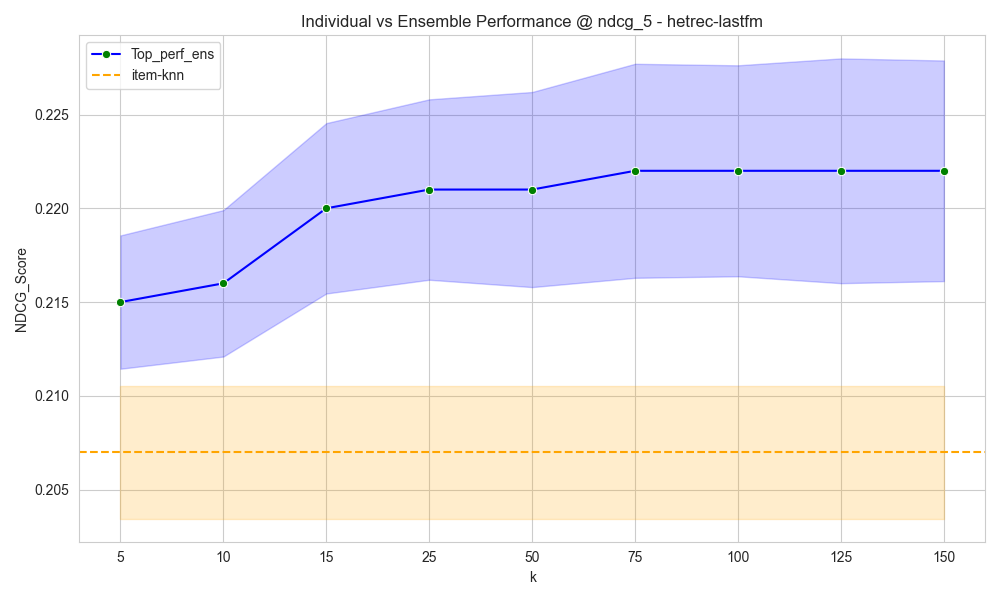} 
    \captionsetup{font=small}
    \caption{hetrec-fm}
    \label{fig:5_hetrec-fm}
  \end{subfigure}
  \hfill
  \begin{subfigure}{0.6\textwidth}
    \includegraphics[width=\linewidth]{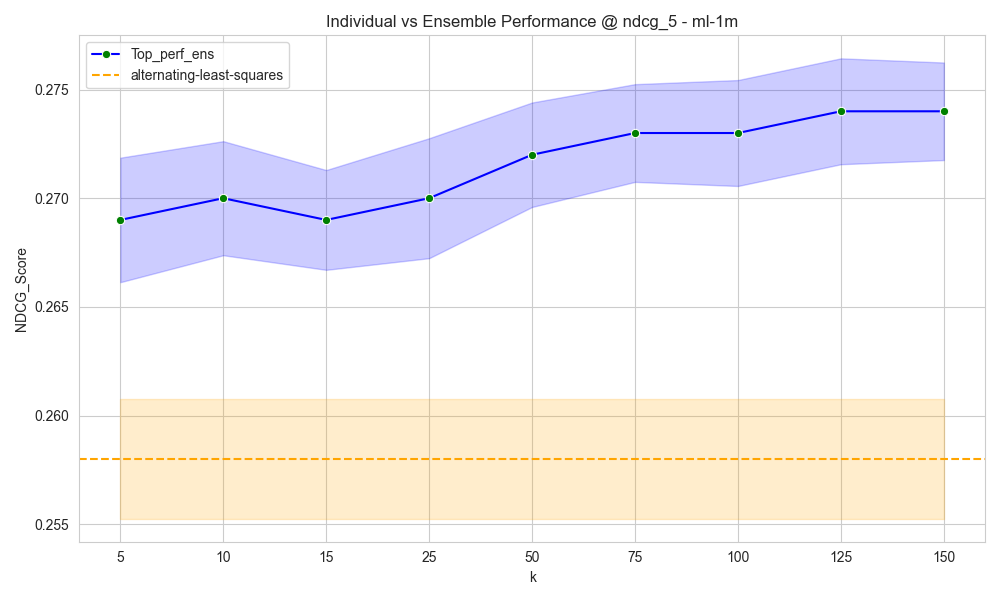}
    \captionsetup{font=small}
    \caption{ml-1m}
    \label{fig:5_ml-1m}
  \end{subfigure}
  \vspace{0.5cm} 
  \hspace{3.5 cm}
  \begin{subfigure}{0.6\textwidth}
    \centering
    \includegraphics[width=\linewidth]{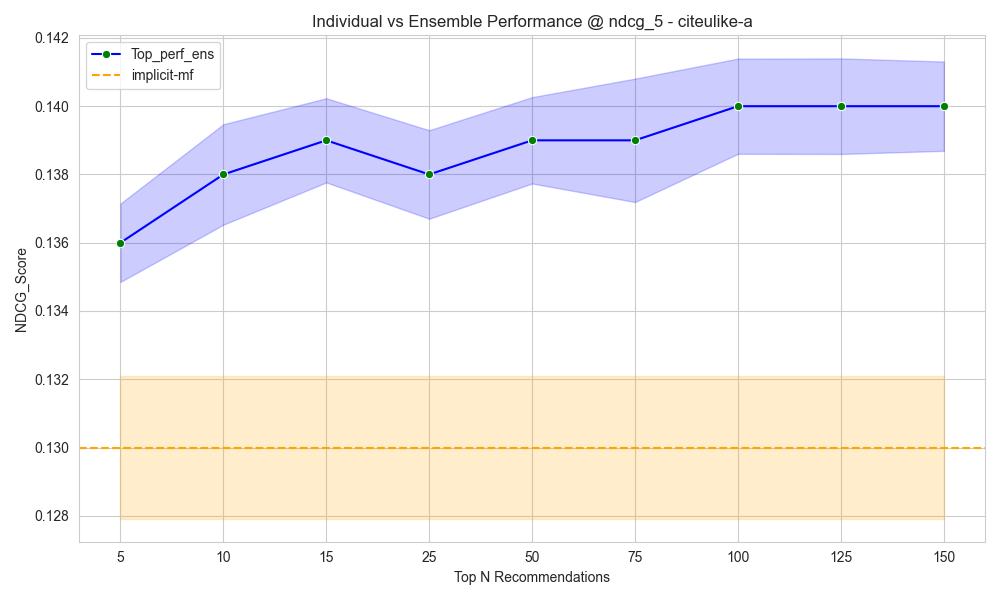}
    \captionsetup{font=small}
    \caption{citeulike-a}
    \label{fig:5_citeulike-a}
  \end{subfigure}
  \caption{"NDCG(@5) Score vs k" across different datasets}
  \label{fig:Ndcg@5}
\end{figure}

\begin{figure}[tbp]
  \begin{subfigure}{0.6\textwidth}
    \includegraphics[width=\linewidth]{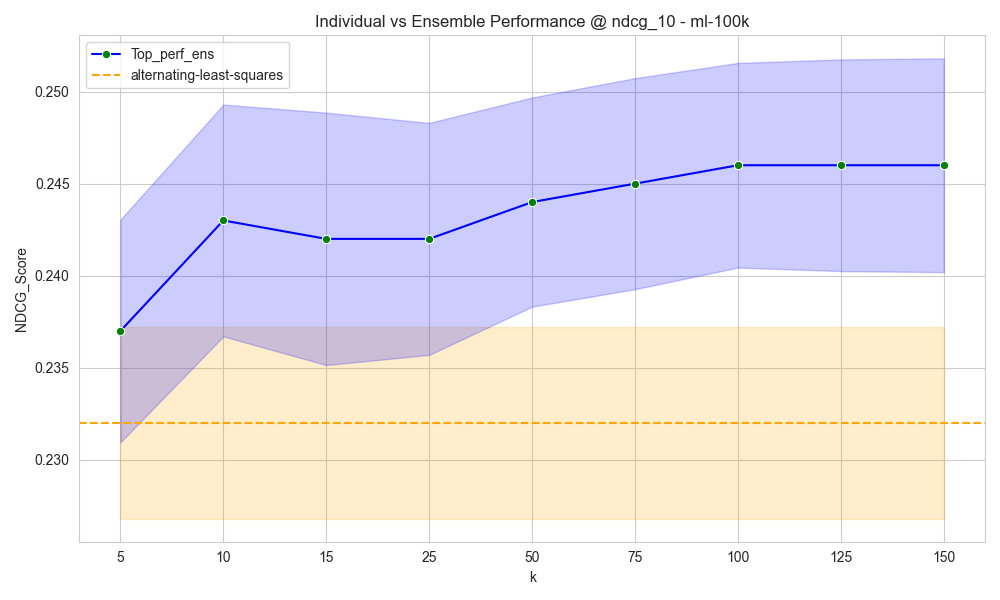} 
    \captionsetup{font=small}
    \caption{ml-100k}
    \label{fig:10_ml-100k}
  \end{subfigure}
  \hfill
  \begin{subfigure}{0.6\textwidth}
    \includegraphics[width=\linewidth]{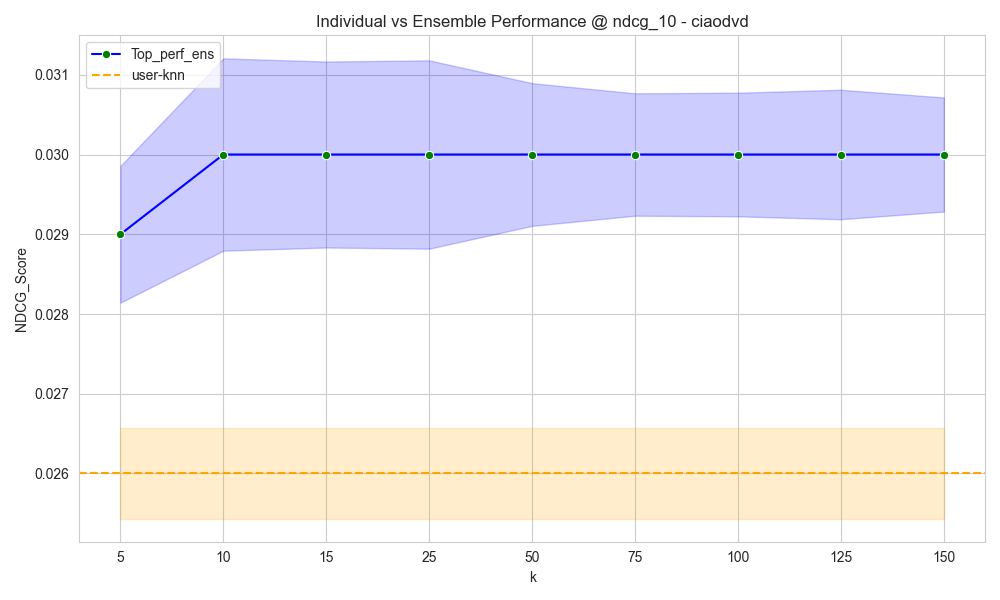} 
    \captionsetup{font=small}
    \caption{ciaodvd}
    \label{fig:10_ciaodvd}
  \end{subfigure}
  \begin{subfigure}{0.6\textwidth}
    \includegraphics[width=\linewidth]{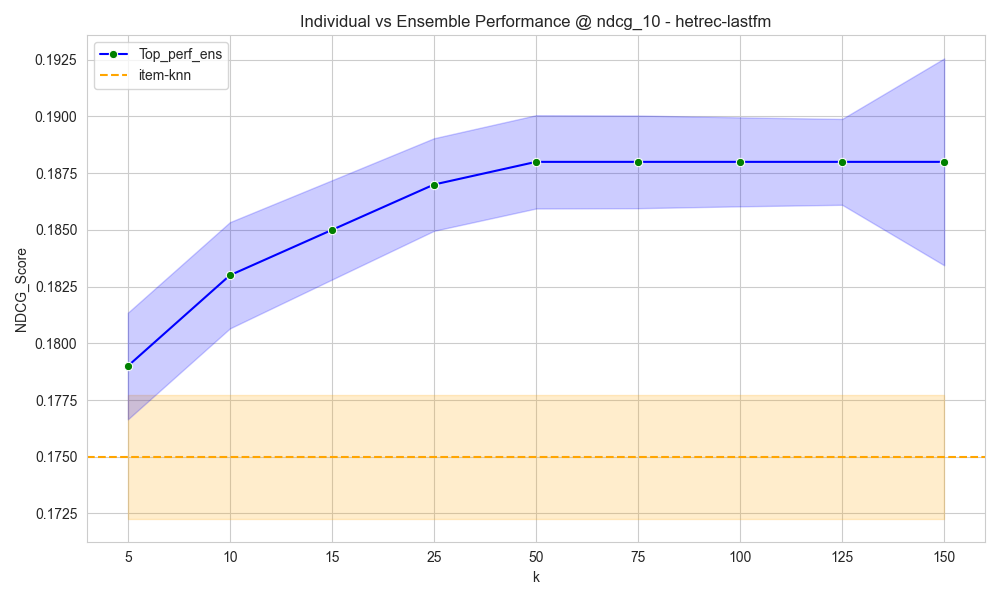} 
    \captionsetup{font=small}
    \caption{hetrec-fm}
    \label{fig:10_hetrec-fm}
  \end{subfigure}
  \hfill
  \begin{subfigure}{0.6\textwidth}
    \includegraphics[width=\linewidth]{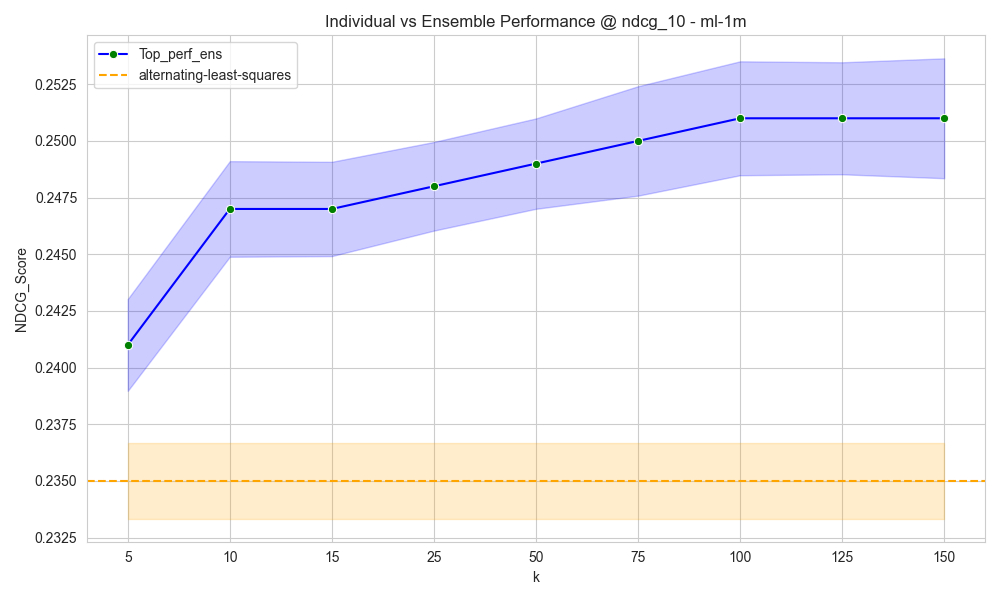}
    \captionsetup{font=small}
    \caption{ml-1m}
    \label{fig:10_ml-1m}
  \end{subfigure}
  \vspace{0.5cm} 
  \hspace{3.5 cm}
  \begin{subfigure}{0.6\textwidth}
    \centering
    \includegraphics[width=\linewidth]{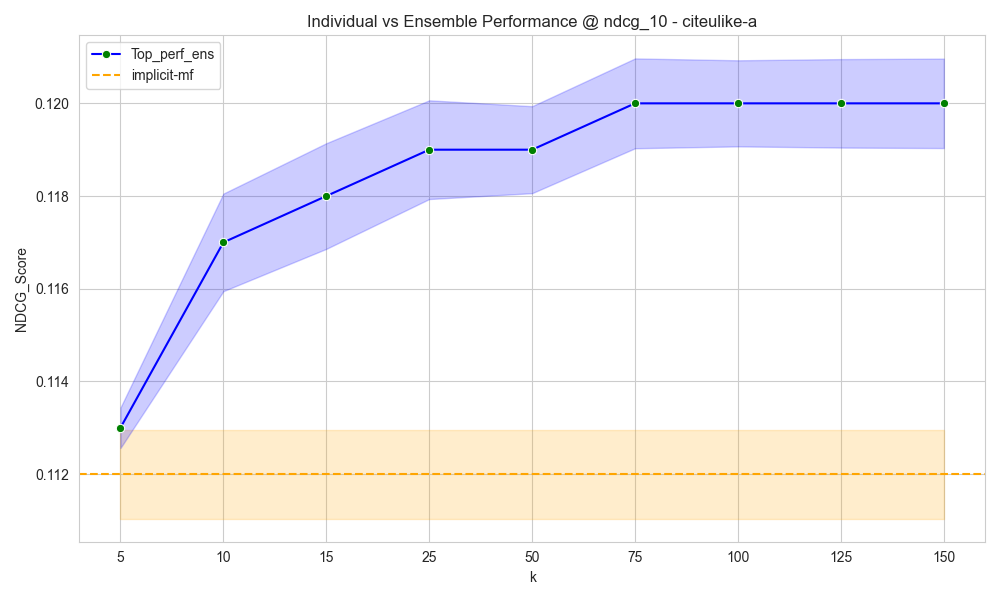}
    \captionsetup{font=small}
    \caption{citeulike-a}
    \label{fig:10_citeulike-a}
  \end{subfigure}
  \caption{"NDCG(@10) Score vs k" across different datasets}
  \label{fig:Ndcg@10}
\end{figure}

\begin{figure}[tbp]
  \begin{subfigure}{0.6\textwidth}
    \includegraphics[width=\linewidth]{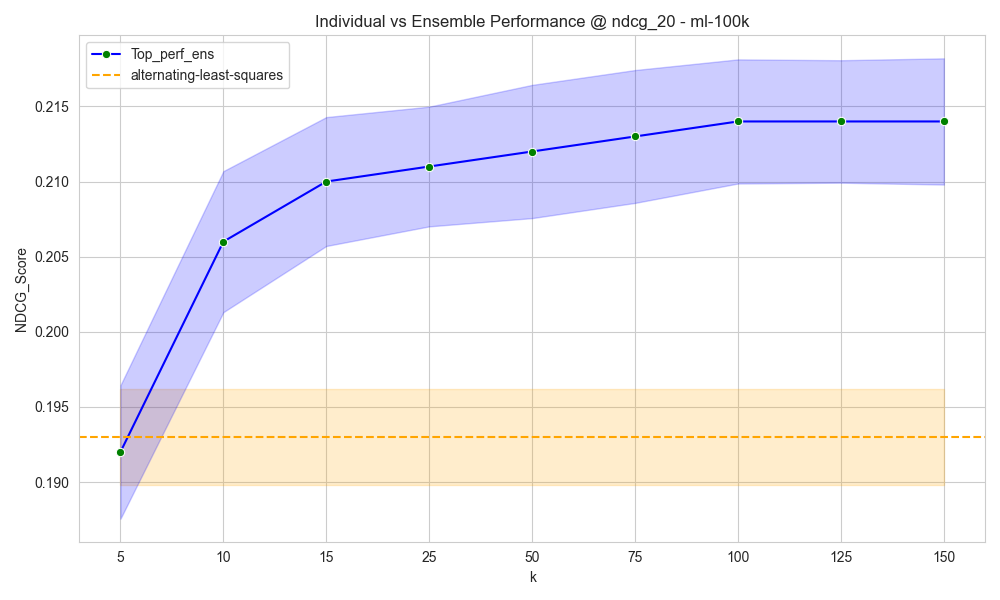} 
    \captionsetup{font=small}
    \caption{ml-100k}
    \label{fig:20_ml-100k}
  \end{subfigure}
  \hfill
  \begin{subfigure}{0.6\textwidth}
    \includegraphics[width=\linewidth]{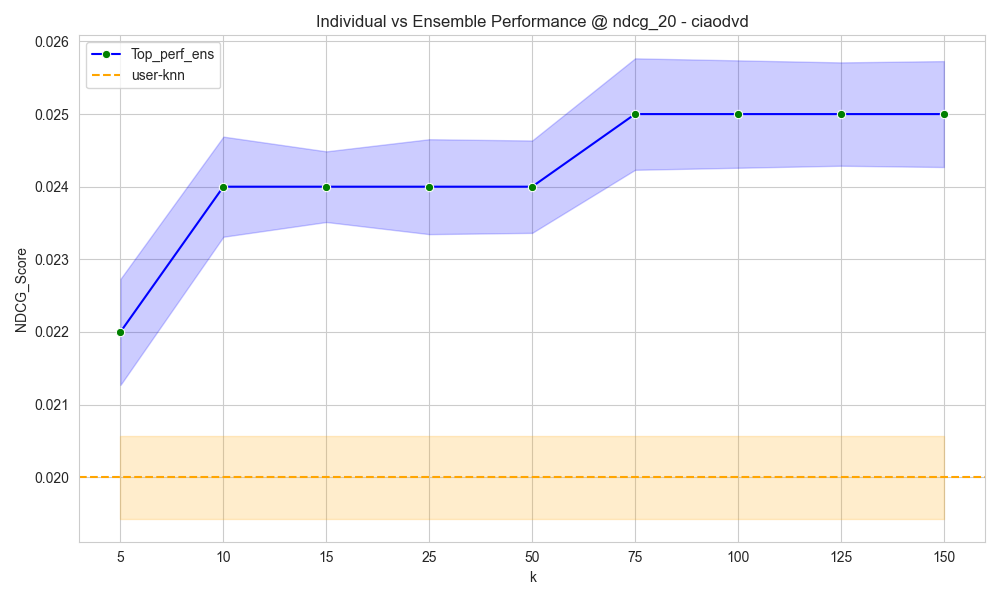} 
    \captionsetup{font=small}
    \caption{ciaodvd}
    \label{fig:20_ciaodvd}
  \end{subfigure}
  \begin{subfigure}{0.6\textwidth}
    \includegraphics[width=\linewidth]{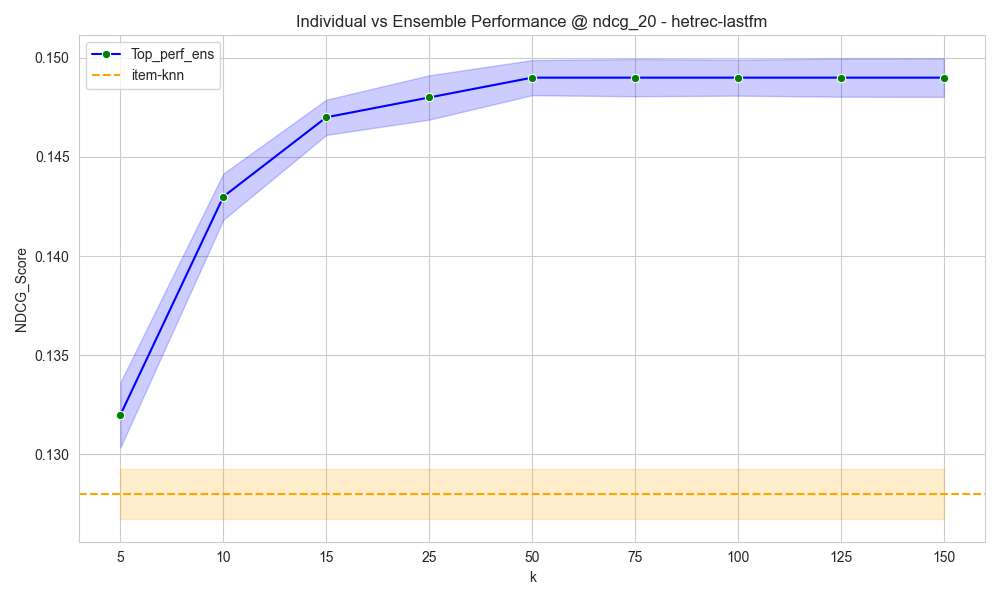} 
    \captionsetup{font=small}
    \caption{hetrec-fm}
    \label{fig:20_hetrec-fm}
  \end{subfigure}
  \hfill
  \begin{subfigure}{0.6\textwidth}
    \includegraphics[width=\linewidth]{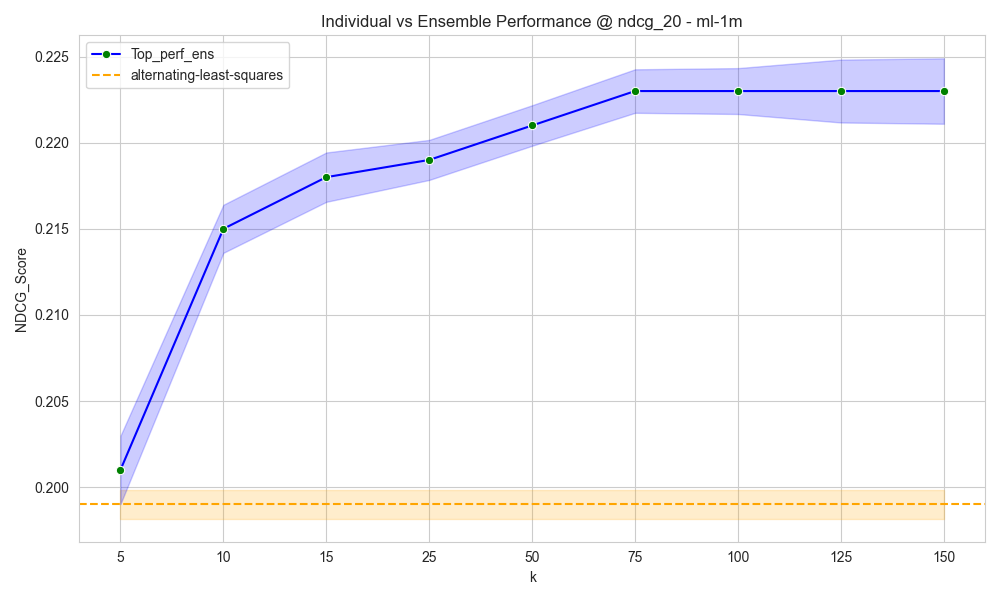}
    \captionsetup{font=small}
    \caption{ml-1m}
    \label{fig:20_ml-1m}
  \end{subfigure}
  \vspace{0.5cm} 
  \hspace{3.5 cm}
  \begin{subfigure}{0.6\textwidth}
    \centering
    \includegraphics[width=\linewidth]{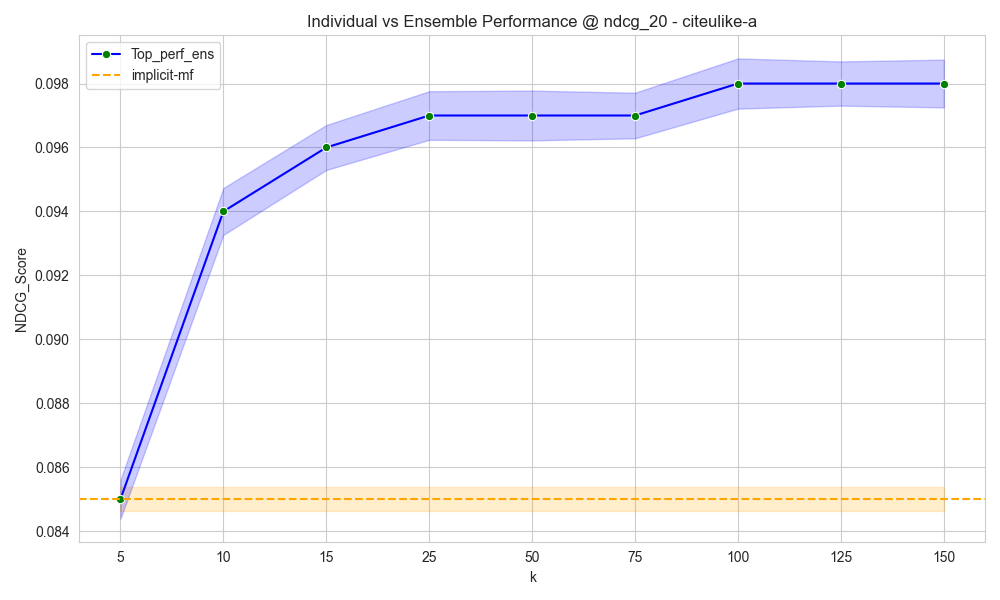}
    \captionsetup{font=small}
    \caption{citeulike-a}
    \label{fig:20_citeulike-a}
  \end{subfigure}
  \caption{"NDCG(@20) Score vs k" across different datasets}
  \label{fig:Ndcg@20}
\end{figure}

Table \ref{NDCG@5} \ref{NDCG@10} \ref{NDCG@20} provides a comparison of the NDCG scores for individual algorithms (optimized specifically for NDCG@5, NDCG@10, and NDCG@20 respectively) and the ensemble method across the five datasets, along with the average performance across all datasets and the relative performance compared to a popularity-based recommender. In the tables, cells highlighted in blue indicate the best-performing individual model for each dataset, while cells highlighted in green represent the best-performing model across all models, including the ensemble method. The Popularity model serves as the baseline. Notably, the Ensemble method consistently outperforms the best individual model across all datasets, demonstrating an average improvement  of 18\% for NDCG@5, 17\% for NDCG@10, and 30\%  for NDCG@20.
\\
\\
In the NDCG@5 table\ref{NDCG@5}, the ensemble technique achieves an impressive performance, outperforming all individual models across the datasets. Particularly noteworthy is its substantial improvement over the popularity-based recommender, with a relative performance of 123\%. This indicates that the ensemble method successfully leverages the strengths of individual models to provide more accurate and relevant recommendations, as evidenced by the significant increase in NDCG scores across all datasets.
\\
\\
Moving on to the NDCG@10 table\ref{NDCG@10}, similar trends are observed, with the ensemble technique consistently surpassing individual models and demonstrating a relative performance of 116\% compared to the popularity-based recommender. This further reinforces the effectiveness of the ensemble approach in enhancing recommendation quality across diverse datasets.
\\
\\
In the NDCG@20 table\ref{NDCG@20}, once again, the ensemble technique stands out as the top performer, showcasing a relative performance of 121\% compared to the popularity-based recommender. This consistent trend across different evaluation metrics and datasets underscores the robustness and reliability of the ensemble technique in improving recommendation quality.
\\
\\
Overall, the results highlight the significant impact of the ensemble technique utilizing the weighted ranking method in enhancing recommendation quality, showcasing its superiority over best individual model across all evaluated metrics and datasets.
\\
\\
The ensemble list of recommendations consists of only top N items, which is derived from top-k list of recommendation of each model. As the ensemble method is based on weighted ranking score considering different values of k might alter the ranking of the items for each, resulting in different top-N list and therefore effecting the NDCG-score. 
\\
\\
For instance, when considering N=5, the top-k items (k=5, 10, 15, 25, 50, 75, 100, 125, and 150) refer to recommendation lists containing k items for a specific user. Each model produces such a list, which is considered during the ensemble creation process. This process generates a list of k items arranged in descending order of the new weighted ranking score for each item(blue line in Figure \ref{fig:Ndcg@5} ). However, since N=5, only the top-5 items are considered, and the NDCG value is evaluated. This value is then compared to the best-performing individual model optimized for NDCG@5(yellow line in Figure \ref{fig:Ndcg@5}), forming the basis for comparison.
\\
\\
In each graph representing NDCG@5, NDCG@10, and NDCG@20, the yellow line corresponds to the NDCG score of the best-performing individual model for the respective NDCG@N across the dataset. Conversely, the blue line illustrates the variation of NDCG scores achieved by the ensemble method as k, the number of recommended items, varies. The shaded band surrounding the lines represents the confidence interval (95\%) for the NDCG scores across the five folds of the dataset, with its size decreasing proportionally with the dataset's magnitude\footnote{Size of the dataset: ml-100k \textless ciaodvd \textless hetrec-fm \textless ml-1m \textless citeulike-a}, indicating higher confidence in the NDCG scores for larger datasets.
\\
\\
For NDCG@5 figure \ref{fig:Ndcg@5}, the graph showcases the fluctuation of NDCG scores across different values of k for each dataset. The comparison between the ensemble's(blue line) and the best-performing individual model's(yellow line) NDCG@5 score reveals that the ensemble outperforms the individual model, highlighting the effectiveness of the ensemble approach in improving recommendation quality.
\\
\\
Similarly, in the case of NDCG@10 figure \ref{fig:Ndcg@10}, a similar trend is observed, with the variation of NDCG scores depicted across different values of k for each dataset. Once again, the comparison between the ensemble and the best-performing individual model underscores instances where the ensemble achieves superior performance, emphasizing its effectiveness in enhancing recommendation quality.
\\
\\
Lastly, for NDCG@20 figure \ref{fig:Ndcg@20}, once again the comparison between the ensemble and the best-performing individual model reveals scenarios where the ensemble outperforms the individual model, further underscoring the effectiveness of the ensemble approach in augmenting recommendation quality.
\\
\\
These detailed analyses provide valuable insights into the performance of the ensemble method across different NDCG values and dataset sizes, showcasing its robustness and superiority over individual models. Through meticulous evaluation and comparison, we highlight the significance of the ensemble approach in optimizing recommendation quality, thus contributing to the advancement of recommender system research and practice.

\section{Conclusion}
This research aimed to explore the efficacy of ensembling techniques in recommender systems, specifically focusing on devising novel ensemble methods to outperform the best-performing individual models, along with conducting a comparative analysis against them. The methodology employed a weighted ranking approach, aggregating different lists of top-N recommendations to generate a final ensemble recommendation list.
\\
\\
Results consistently indicate that the ensemble method surpasses the performance of individual models across all datasets. Moreover, it was observed that the confidence band of the ensemble method decreases with an increase in the dataset size. These findings underscore the potential of ensemble techniques to enhance recommendation systems, similar to their effectiveness in regression and classification tasks.
\\
\\
The research highlights the importance of applying ensembling in recommendation systems and its ability to improve results. However, it also acknowledges the computational intensity of the weighted ranking method, which necessitates significant time for experimentation.
\\
\\
Our study represents an important step forward in the application of ensemble techniques to recommender systems. By demonstrating the superiority of ensembles over individual models, we lay the groundwork for future research endeavors aimed at enhancing recommendation systems. Our findings underscore the transformative potential of ensemble techniques in reshaping recommendation landscapes, providing users with recommendations that align closely with their preferences and needs.

\bibliography{bibliography}{}
\bibliographystyle{plain}
\end{document}